\documentclass[prd,twocolumn,showpacs,preprintnumbers,nofootinbib,amsmath,amssymb]{revtex4}

\usepackage{graphicx}
\usepackage{dcolumn}
\usepackage{bm}
\usepackage{amsmath}

\newcommand{\beq}{\begin{eqnarray}}
\newcommand{\eeq}{\end{eqnarray}}

\newcommand{\real}{{\sf I}\kern-.12em{\sf R}}
\newcommand{\comp}{{\sf I}\kern-.50em{\sf C}}
\newcommand{\unity}{{\sf I}\kern-.54em{\sf 1}}

\def\spose#1{\hbox to 0pt{#1\hss}}
\def\ltapprox{\mathrel{\spose{\lower 3pt\hbox{$\mathchar"218$}}
 \raise 2.0pt\hbox{$\mathchar"13C$}}}

\begin{document}

\title{The critical line from imaginary to real baryonic chemical
  potentials in two-color QCD}
\author{Paolo Cea}
\affiliation{
Dipartimento di Fisica dell'Universit\`a di Bari and INFN - Sezione di Bari, I-70126 Bari, Italy}
\email{paolo.cea@ba.infn.it}
\author{Leonardo Cosmai}
\affiliation{INFN - Sezione di Bari, I-70126 Bari, Italy}
\email{leonardo.cosmai@ba.infn.it}
\author{Massimo D'Elia}
\affiliation{Dipartimento di Fisica dell'Universit\`a di Genova and INFN - Sezione di Genova, I-16146 Genova, Italy}
\email{delia@ge.infn.it}
\author{Alessandro Papa}
\affiliation{
Dipartimento di Fisica dell'Universit\`a della Calabria
and INFN - Gruppo collegato di Cosenza,
I-87036 Arcavacata di Rende, Cosenza, Italy}
\email{papa@cs.infn.it}

\date{\today}

\begin{abstract}
The method of analytic continuation from imaginary to real chemical
potentials $\mu$ is one of the few available techniques to study QCD at
finite temperature and baryon density. One of its most appealing applications
is the determination of the critical line for small $\mu$:
we perform a direct test of the validity of the method
in this case by studying two-color QCD, where the sign problem is absent.
The (pseudo)critical line is found to be analytic around $\mu^2 = 0$, but a very large precision
would be needed at imaginary $\mu$ to correctly predict the
location of the critical line at real $\mu$.
\end{abstract}


\maketitle

\section{Introduction}
\label{introd}

The study of strong interactions at finite temperature and
baryon density is
relevant to fundamental phenomenological
issues, like the experimental search for the deconfinement
transition in heavy ion collisions or the properties of compact
astrophysical objects.
Several questions need to be clarified: the structure of the QCD
phase diagram, i.e. the location and the nature of the
transition lines, as well as the properties of strongly interacting
matter, specially close to the transition lines.
Unfortunately numerical lattice simulations, which are
the ideal non-perturbative tool for first principle theoretical
studies, are hindered in presence of a baryon chemical potential
by the fermionic determinant being complex, which makes
importance sampling techniques not usable.

Various possibilities
have been explored to circumvent the problem:
reweighting techniques~\cite{glasgow,fodor,density},
the use of an imaginary chemical potential either
for analytic
continuation~\cite{muim,immu_dl,azcoiti,chen,giudice,potts3d,cea,defor06,sqgp,conradi,cea1,kt}
or for reconstructing the canonical
partition function~\cite{rw,cano1,cano2}, Taylor expansion
techniques~\cite{taylor1,taylor2} and
non-relativistic expansions~\cite{hmass1,hmass2,hmass3}.
All these methods do not really solve the problem and are expected
to work well in a limited range of parameters. Cross-checks among
them and direct test of the methods in simpler models are therefore
extremely important.

In the present paper we consider the method of analytic continuation.
The fermionic determinant, which is complex in presence of
a real baryonic chemical potential $\mu$,
is instead real if $\mu$ is purely imaginary, making usual
Monte Carlo simulations feasible.
Due to CP invariance, the partition function $Z$ is an even function of $\mu$,
$Z = Z(T,\mu^2)$.
Numerical results obtained at imaginary $\mu$'s
($\mu^2 < 0$) can be used to fit the behaviour of physical observables
with suitable functions, which can then be continued to $\mu^2 > 0$.
The method is expected to fail outside
analyticity domains, hence in particular when crossing physical or
unphysical transition lines; but numerical instabilities, strictly
linked to the choice of interpolating functions, can be source of
systematic errors also elsewhere. A model where numerical simulations
are possible both at imaginary and real values of $\mu$ is
the ideal testground to study such systematic effects and the general
reliability of the method: an example is provided by two-color
QCD, where various studies regarding the analytic continuation of
physical observables have been performed in the past~\cite{giudice,cea,cea1}.

One of the most important applications of analytic continuation is
the determination of the critical line or of critical surfaces
for small values of $\mu$~\cite{muim,immu_dl,azcoiti,chen,defor06}.
The theoretical basis in this case is
not as straightforward as for physical observables and is based on
the assumption that susceptibilities, whose peaks signal the presence
of the transition, be analytic functions of the parameters on a finite
volume~\cite{muim}. Direct tests of the method are
even more important in this case: therefore we
extend our analysis of two-color QCD to the study of the
(pseudo)critical line.
As in usual QCD simulations,
we will determine locations of the critical line for
$\mu^2 < 0$ and interpolate them by
suitable functions to be continued to $\mu^2 > 0$. The prediction
obtained at real $\mu$ will then be compared with direct
determinations of the transition line.

\section{Numerical results}
\label{numres}

As in Ref.~\cite{cea}, to which we refer for further details about our
numerical setup, we have performed numerical simulations on a
$16^3\times 4$ lattice of the SU(2) gauge theory with $n_f=8$ degenerate
staggered fermions having mass $am=0.07$. The algorithm adopted has been the usual
exact $\phi$ algorithm described in Ref.~\cite{Gottlieb:1987mq},
properly modified for the inclusion of a finite chemical potential.
Simulations have been performed on the APE100 and APEmille crates in Bari and
on the computer facilities at the INFN apeNEXT Computing Center in Rome.

\begin{figure}[t!]
\vspace{-0.05cm}
\includegraphics*[height=0.295\textheight,width=0.95\columnwidth]{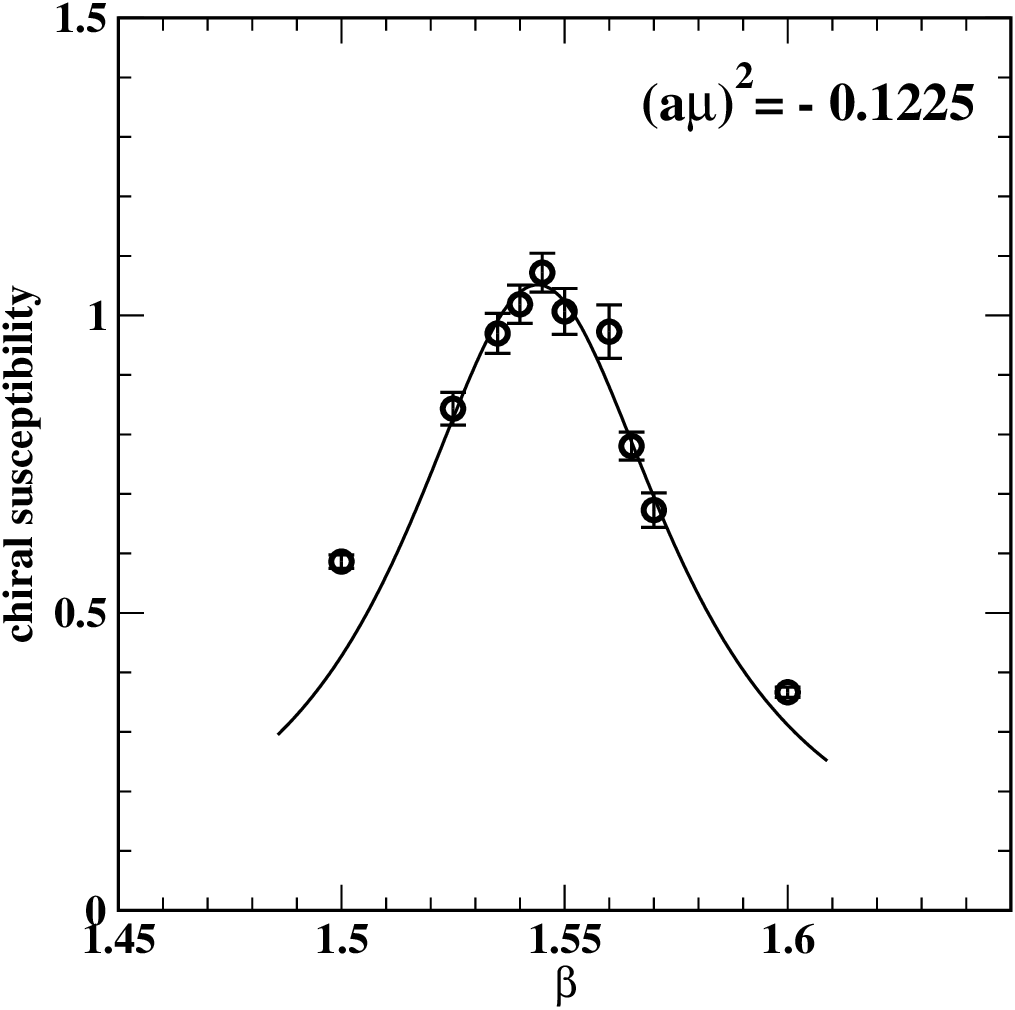}
\vspace{-0.4cm}
\caption{Chiral susceptibility at $(a\mu)^2 = -0.1225$ {\it vs} $\beta$.}
\label{fig1}
\vspace{-0.2cm}
\end{figure}

\begin{figure}[t!]
\vspace{-0.1cm}
\includegraphics*[height=0.295\textheight,width=0.95\columnwidth]{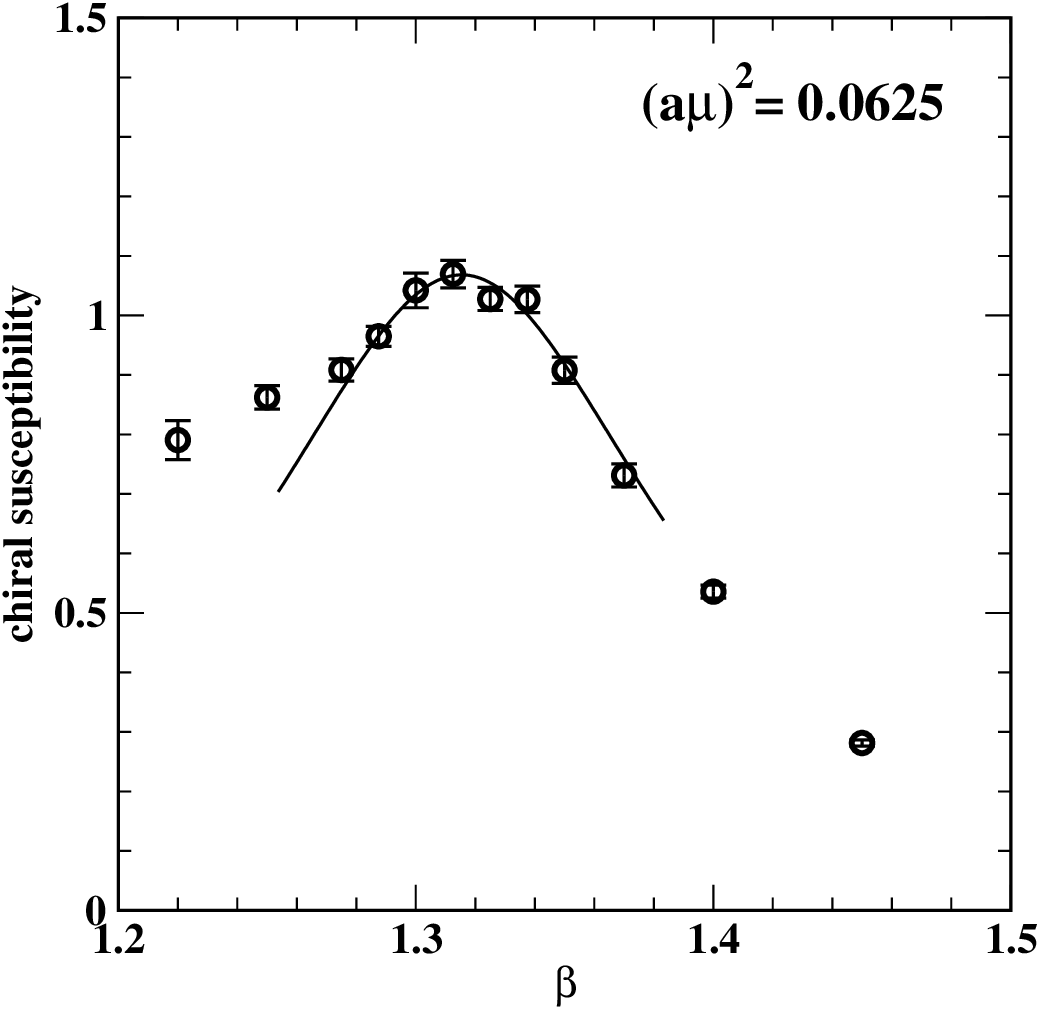}
\vspace{-0.4cm}
\caption{Chiral susceptibility at $(a\mu)^2 = 0.0625$ {\it vs} $\beta$.}
\label{fig2}
\vspace{-0.2cm}
\end{figure}

In absence of true non-analyticities at the transition line, as
on a finite volume, the location of the critical line
may be dependent on the observable chosen to probe the transition. For
that reason we have decided to determine, for a set of $\mu^2$ values,
the critical couplings $\beta_c (\mu^2)$ by looking at the
peaks of the susceptibilities of three different observables: the
chiral condensate, the Polyakov loop and the plaquette. We have taken
the $\mu^2$ values on the imaginary side in the so-called
first Roberge-Weiss (RW) sector, i.e. in the range delimited by
$(a\mu)^2=-(\pi/8)^2$ and $\mu^2=0$ (see Ref.~\cite{cea} for a detailed
discussion on the phase diagram in the $T - i \mu$ plane).

In Figs.~\ref{fig1} and \ref{fig2} we show as examples the behavior
with $\beta$ of the chiral susceptibility at $(a\mu)^2=-0.1225$ and
$(a\mu)^2=0.0625$, respectively. In Table~\ref{tab1} we summarize
our determinations of $\beta_c(\mu^2)$, obtained by fitting the
peaks of the susceptibilities according to a Lorentzian function.
The values of
$\beta_c(\mu^2)$ depend very weakly on the ``probe'' observable;
only in one case the relative deviation between two determinations
at the same $\mu^2$ slightly exceeds 3$\sigma$.

We have tried at first to determine the critical line by interpolating
the values of $\beta_c(\mu^2)$ for $\mu^2 \leq 0$ only, i.e. for the
first seven entries in Table~\ref{tab1}, and have repeated this procedure for
each of the three observables considered.

\begin{table*}[t]
\setlength{\tabcolsep}{0.5pc}
\centering
\caption[]{Summary of the critical values of $\beta$ obtained by fitting
the peaks of the susceptibilities of chiral condensate $\langle\overline\psi\psi\rangle$,
Polyakov loop $\langle L \rangle$ and plaquette $\langle P \rangle$. The value
of $\chi^2/{\rm d.o.f.}$ is given in each case.}
\begin{tabular}{ddcdcdc}
\hline
\hline
\multicolumn{1}{c}{\hspace{0.70cm}$(a\mu)^2$} &
\multicolumn{1}{c}{\hspace{1cm}$\langle\overline\psi\psi\rangle$} &
$\chi^2/{\rm d.o.f.}$ &
\multicolumn{1}{c}{\hspace{1cm}$\langle L \rangle$} &
$\chi^2/{\rm d.o.f.}$ &
\multicolumn{1}{c}{\hspace{1cm}$\langle P \rangle$} &
$\chi^2/{\rm d.o.f.}$ \\
\hline
-0.1225 & 1.5440(16) & 1.34 & 1.5349(43) & 0.85 & 1.5418(24) & 0.93 \\
-0.09   & 1.5068(15) & 0.65 & 1.5019(29) & 0.25 & 1.5046(21) & 1.06 \\
-0.0625 & 1.4775(29) & 0.88 & 1.4665(32) & 0.31 & 1.4755(37) & 0.65 \\
-0.04   & 1.4532(16) & 0.50 & 1.4453(36) & 0.76 & 1.4522(26) & 1.21 \\
-0.0225 & 1.4324(22) & 1.20 & 1.4240(28) & 0.80 & 1.4300(39) & 0.80 \\
-0.01   & 1.4197(16) & 1.86 & 1.4104(33) & 0.43 & 1.4199(26) & 1.45 \\
 0.     & 1.4102(18) & 0.07 & 1.3989(61) & 0.49 & 1.4117(32) & 0.07 \\
 0.04   & 1.3528(22) & 2.91 & 1.3388(72) & 1.01 & 1.3631(46) & 1.16 \\
 0.0625 & 1.3145(30) & 1.34 & 1.2976(62) & 0.87 & 1.3286(50) & 1.28 \\
 0.09   & 1.2433(59) & 1.09 & 1.2508(62) & 0.98 & 1.2864(109)& 0.60 \\
\hline
\hline
\end{tabular}
\label{tab1}
\end{table*}

In all cases we have found that the optimal interpolating function is a polynomial
of the form $A+B(a\mu)^2$. If different functions are used, such as larger order polynomials
or ratio of polynomials, the fit puts to values compatible with zero all parameters
except two of them, thus reducing again to a first order polynomial in $\mu^2$.
Moreover, the uncertainty on the parameters turns to be so large to make the extrapolation to
$\mu^2>0$ useless. This is in marked difference with what we found in Ref.~\cite{cea}
for the analytic continuation of physical observables, where ratio of polynomials performed rather well.

\begin{figure}[t!]
\vspace{-0.05cm}
\includegraphics*[height=0.295\textheight,width=0.95\columnwidth]{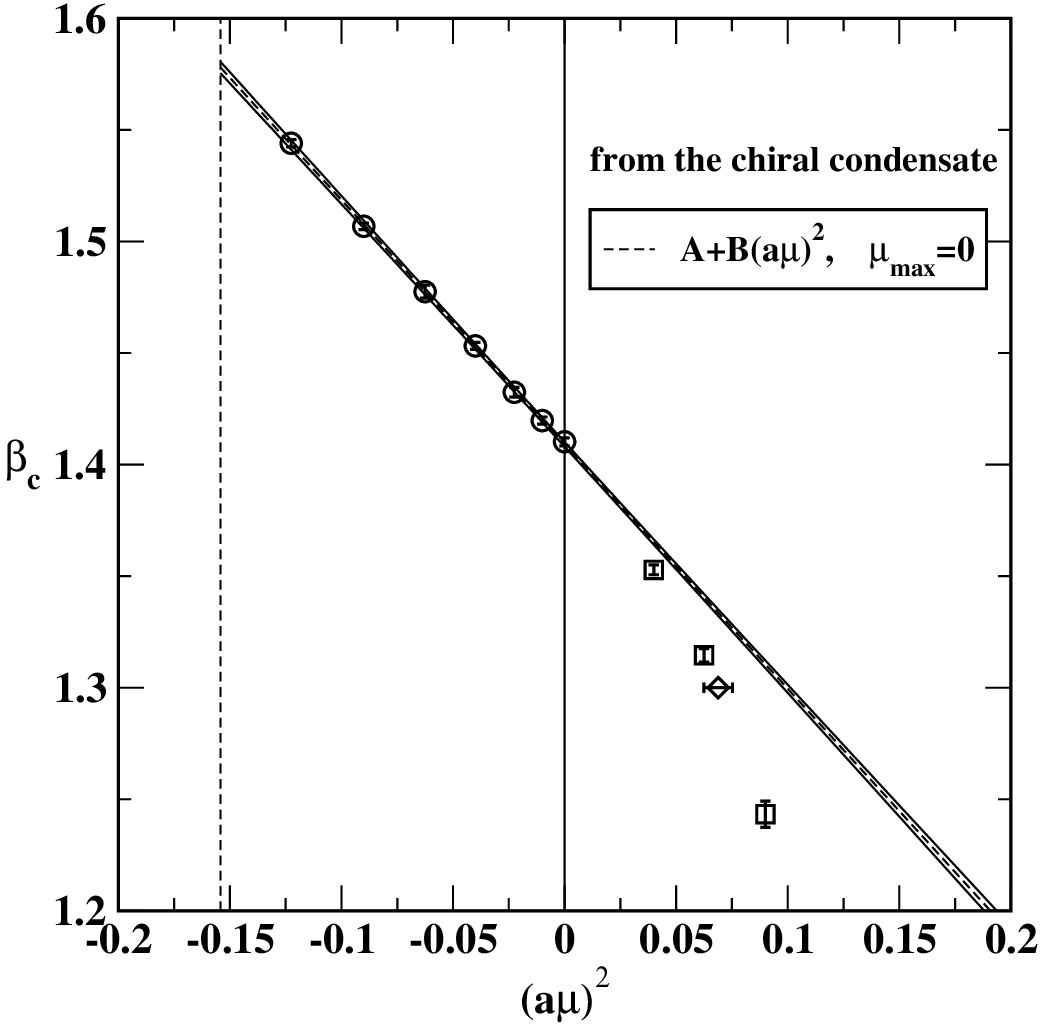}
\vspace{-0.3cm}
\caption{
Critical couplings obtained from the chiral susceptibility, together
with a
linear fit in $(a\mu)^2$ to data with $\mu^2 \leq 0$.
The datum at
$\beta_c=1.30$ (diamond) is taken from Ref.~\cite{cea}.
}
\label{fig:crit_psibpsi}
\vspace{-0.2cm}
\end{figure}

The fit results are summarized in Table~\ref{tab2} (rows for which the last entry is
$(a\mu_{\rm max})^2=0$). The fit parameters have a tiny dependence on
the observable considered. Moreover, the extrapolation of the critical line at $(a\mu)^2=-(\pi/8)^2$,
corresponding to the first RW transition, is in good agreement with an independent
determination of the RW endpoint~\cite{Cor07}. This is a clear indication that
the critical line is a continuation of the high-temperature first
order RW transition line at $(a\mu)^2=-(\pi/8)^2$.

\begin{figure}[t!]
\vspace{-0.2cm}
\includegraphics*[height=0.295\textheight,width=0.95\columnwidth]{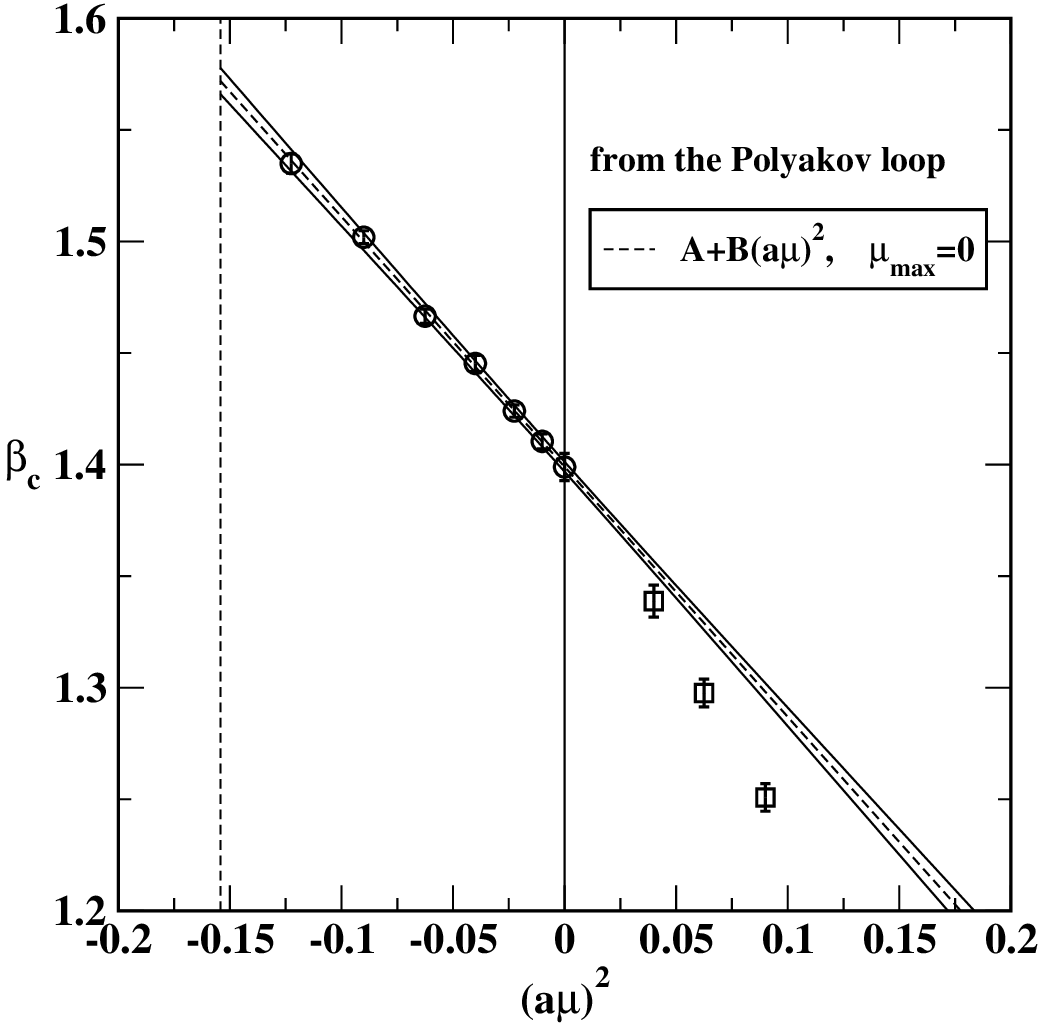}
\vspace{-0.3cm}
\caption{
Same as in Fig.~\ref{fig:crit_psibpsi} for the critical values of
$\beta$ obtained via the Polyakov loop susceptibility.
}
\label{fig:crit_poly}
\vspace{-0.2cm}
\end{figure}

\begin{figure}[t!]
\vspace{-0.2cm}
\includegraphics*[height=0.295\textheight,width=0.95\columnwidth]{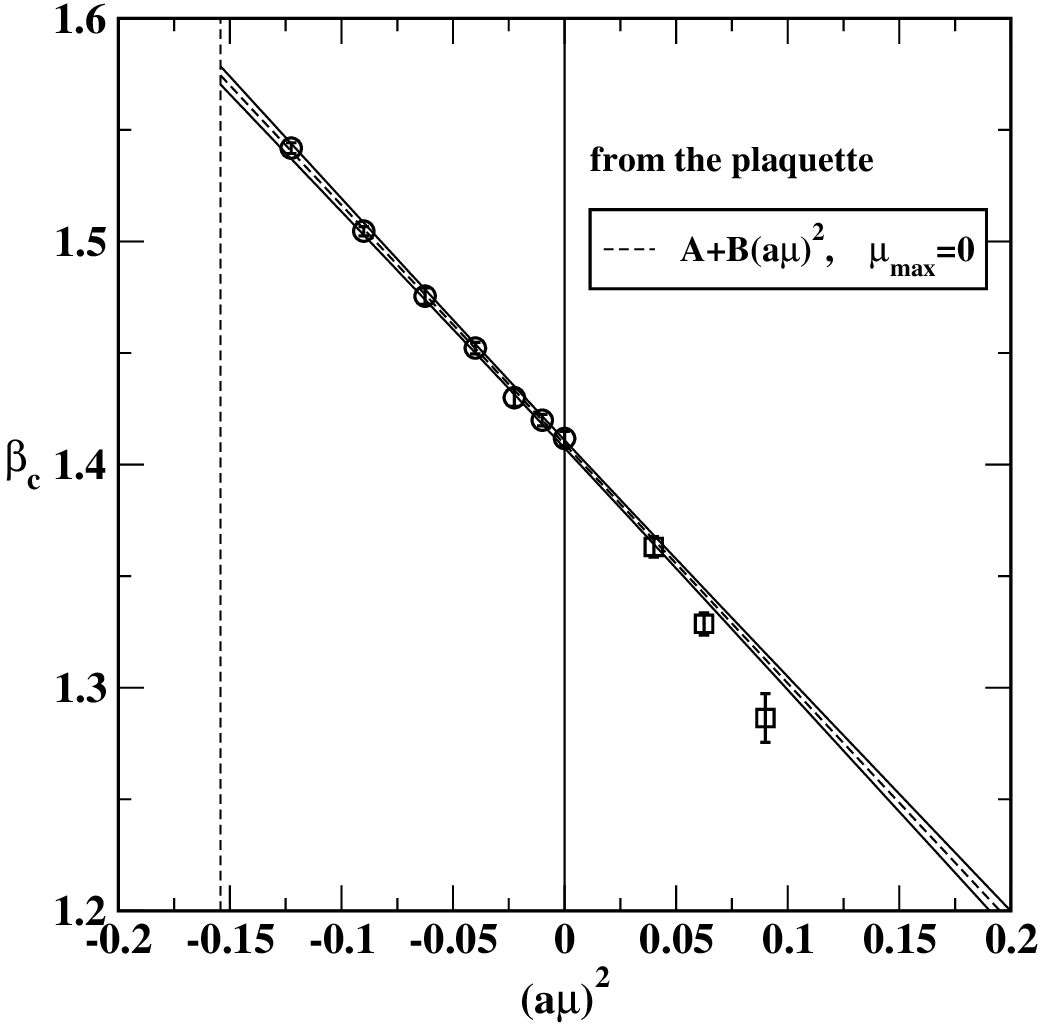}
\vspace{-0.3cm}
\caption{
Same as in Fig.~\ref{fig:crit_psibpsi} for the critical values of
$\beta$ obtained via the plaquette susceptibility.
}
\label{fig:crit_plaq}
\vspace{-0.2cm}
\end{figure}

The extrapolation of the critical line to $\mu^2>0$ can be compared with the direct determinations of
$\beta_c$ at $(a\mu)^2=0.04$, 0.0625 and 0.09. As one can see from Figs.~\ref{fig:crit_psibpsi},
\ref{fig:crit_poly} and \ref{fig:crit_plaq}, there is discrepancy between the extrapolated critical line
and the direct determinations of $\beta_c(\mu^2)$ for each of the observables considered. The
discrepancy is more pronounced when the susceptibility of the chiral condensate is used.
This result implies that either the critical line is not analytic
on the whole interval of $\mu^2$ values considered or it is in fact
analytic, but the interpolation at $\mu^2\leq 0$ is not accurate enough
to correctly reproduce the behavior at $\mu^2>0$.

\begin{table*}[t]
\setlength{\tabcolsep}{0.4pc}
\centering
\caption[]{Parameters of various fits
of the critical couplings according to
$\beta_c (\mu^2) = A + B (a\mu)^2 + C (a\mu)^4 + D (a\mu)^6$ (polynomial)
or $\beta_c (\mu^2) = [A + B (a\mu)^2]/[1 + C (a\mu)^2]$ (ratio).
Blank columns stand for terms not included
in the fit.
The last column reports the largest value of $(a\mu)^2$
included in each fit.}
\begin{tabular}{ccddddcc}
\hline
\hline
observable & fit & \multicolumn{1}{c}{\hspace{1cm}$A$} & \multicolumn{1}{c}{\hspace{1cm}$B$}
& \multicolumn{1}{c}{\hspace{1cm}$C$} & \multicolumn{1}{c}{\hspace{1cm}$D$} &
$\chi^2$/d.o.f. & $(a\mu_{\rm max})^2$ \\
\hline
$\langle\overline\psi\psi\rangle$ & polynomial & 1.4091(10) & -1.095(15) &  &  & 0.27 & 0.00 \\
$\langle L \rangle$               & polynomial & 1.3990(22) & -1.121(35) &  &  & 0.28 & 0.00 \\
$\langle P \rangle$               & polynomial & 1.4092(17) & -1.072(24) &  &  & 0.38 & 0.00 \\
\hline
$\langle\overline\psi\psi\rangle$ & polynomial & 1.4096(13) & -1.055(60) &  0.33(48) & & 1.35 & 0.00 \\
$\langle L \rangle$               & polynomial & 1.3989(32) & -1.13(13)  & -0.1(1.1) & & 0.35 & 0.00 \\
$\langle P \rangle$               & polynomial & 1.4105(22) & -0.986(94) &  0.69(73) & & 0.26 & 0.00 \\
\hline
$\langle\overline\psi\psi\rangle$ & ratio      & 1.4096(13) & -0.64(65)  &  0.30(42) & & 0.22 & 0.00 \\
$\langle L \rangle$               & ratio      & 1.3989(32) & -1.22(1.45)& -0.06(95) & & 0.35 & 0.00 \\
$\langle P \rangle$               & ratio      & 1.4105(22) & -0.12(97)  &  0.62(63) & & 0.26 & 0.00 \\
\hline
$\langle\overline\psi\psi\rangle$ & polynomial & 1.4088(99) & -1.230(25) & -3.77(35) & -22.7(3.6) & 1.00 & 0.09 \\
$\langle L \rangle$               & polynomial & 1.3963(22) & -1.303(53) & -2.71(52) & -11.2(6.6) & 0.43 & 0.09 \\
$\langle P \rangle$               & polynomial & 1.4096(17) & -1.110(40) & -1.78(60) & -12.6(5.7) & 0.29 & 0.09 \\
\hline
$\langle\overline\psi\psi\rangle$ & ratio      & 1.4062(87) & -3.79(24)  & -1.75(17) &            & 5.04 & 0.09 \\
$\langle L \rangle$               & ratio      & 1.3948(18) & -3.67(39)  & -1.66(27) &            & 0.58 & 0.09 \\
$\langle P \rangle$               & ratio      & 1.4075(14) & -2.07(44)  & -0.64(30) &            & 0.90 & 0.09 \\
\hline
\hline
\end{tabular}
\label{tab2}
\end{table*}

\begin{figure}[t!]
\vspace{-0.05cm}
\includegraphics*[height=0.295\textheight,width=0.95\columnwidth]{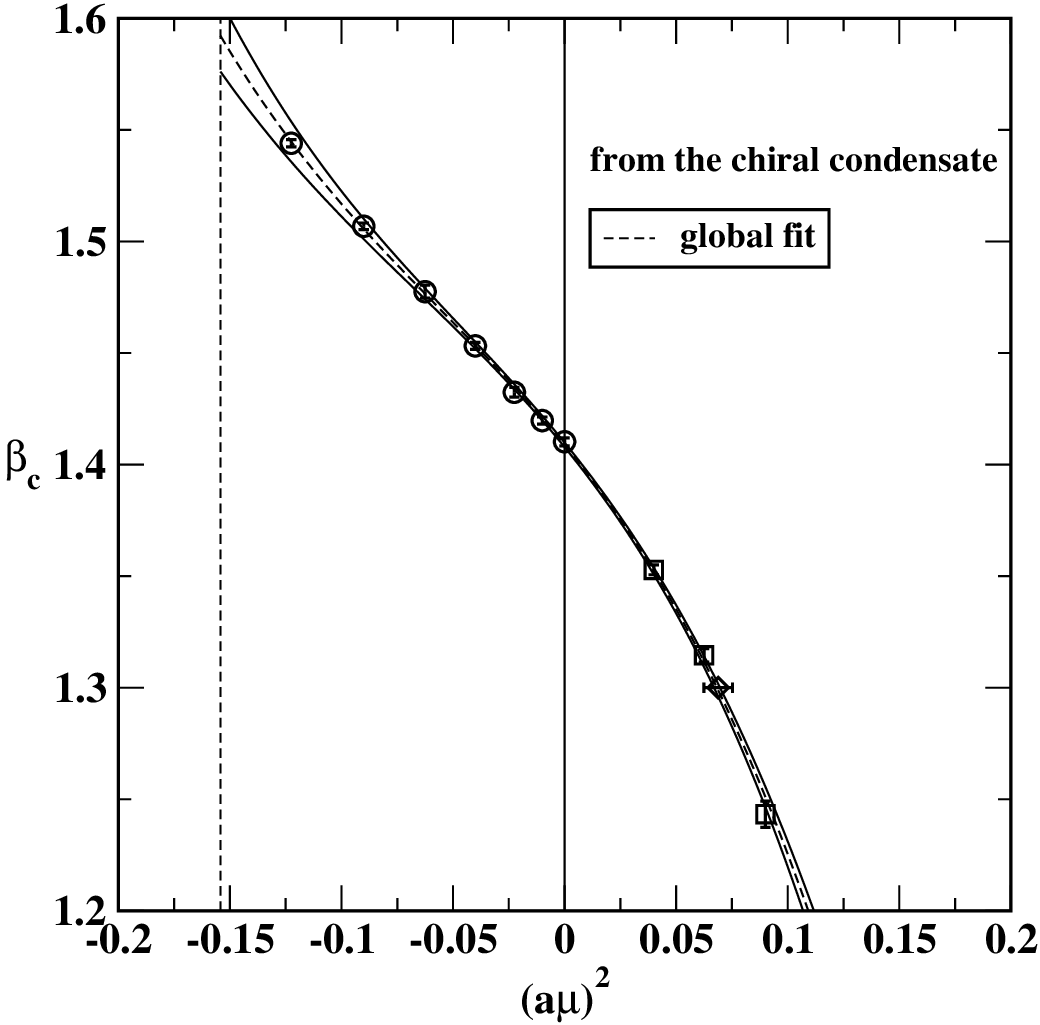}
\vspace{-0.3cm}
\caption{
Same as in Fig.~\ref{fig:crit_psibpsi}, but with the results of a fit to all data
including terms up to $\mu^6$.
}
\label{fig:crit_psibpsi_all}
\vspace{-0.2cm}
\end{figure}

In order to discriminate between these two possibilities, we have
interpolated all the available determinations of $\beta_c(\mu^2)$,
i.e. data for both $\mu^2\leq 0$ and $\mu^2>0$, with a unique analytic
function. We have found that a polynomial of third order in $\mu^2$
nicely fits all data for $\beta_c(\mu^2)$ for each of the three
``probe'' observables -- see Fig.~\ref{fig:crit_psibpsi_all} for the
case of the chiral condensate and Table~\ref{tab2} for a summary of the
fit parameters and of the $\chi^2$/d.o.f.'s (rows for which the last entry is
$(a\mu_{\rm max})^2=0.09$).

What happened here is a sort of ``conspiracy'': the $\mu^4$ and
$\mu^6$ terms compensate each other at large negative values of $\mu^2$
so that the effective interpolating function of the data at $\mu^2\leq 0$ is
a first order polynomial in $\mu^2$. For $\mu^2>0$, instead, the $\mu^4$ and
$\mu^6$ terms work in the same direction and their contribution cannot be neglected.

For the cases of Polyakov look and plaquette, but not for the chiral condensate,
we have obtained nice global interpolations also by means of a ratio of
polynomials of the form $(A+B(a\mu)^2)/(1+C (a\mu)^2)$.

\section{Discussion and conclusions}
\label{concl}

In this paper, we have tested in two-color QCD the analytic continuation of the critical
line in the $T -  \mu$ plane from imaginary to real chemical potential.
We have found that the critical line around $\mu=0$ can be described by an analytic function.
Indeed, a third order polynomial in $\mu^2$ nicely fits all the available data for the
critical coupling. However, when trying to infer the behavior of
the critical line at real $\mu$ from the extrapolation of its behavior at imaginary
$\mu$, a very large precision would be needed to get the correct
result.
In the case of polynomial interpolations
there is a clear indication that high-order terms
play a relevant role at $\mu^2 > 0$ but are less visible at $\mu^2 <
0$,  this calling for an accurate knowledge of the critical line in all the
first RW sector. This scenario could be peculiar of two-color QCD. If
confirmed in other theories free of the sign problem, such as QCD at finite isospin density,
then one should seriously reconsider the analytic continuation of the critical line
in the physically relevant case of QCD at finite baryon density.
There the critical line at imaginary $\mu$
has been interpolated in most cases by a first order polynomial in $\mu^2$, the only exception being
the use of Pad\'e approximants in Ref.~\cite{lom}. If the scenario described in this paper,
and in particular the ``conspiracy'' between the $\mu^4$ and $\mu^6$ terms, is robust, then
it may be useful to revisit the interpolations used in QCD and to make the effort of
determining accurately at least one more term in the polynomial fit.

\end{document}